\documentstyle[12pt,leqno]{article}
\begin{document}
\title{QUANTUM MECHANICS IN THE NONCONTRACTIBLE SPACE}
\author{D. PALLE \\Department of Theoretical Physics,Rugjer Bo\v skovi\'c 
Institute \\P.O.Box 1016,Zagreb,CROATIA}
\date{ }
\maketitle
{\bf Summary.} - 
We show that the impact of the fundamental length in quantum mechanics 
can be studied within the formalism of Berry's geometrical phase
with the line broadening as a resulting physical effect.
\hspace*{12 mm}03.65 Bz Foundations,
 theory of measurement, miscellaneous theories  \\
\hspace*{29 mm} (including Aharonov-Bohm effect, 
 Bell inequalities, \\ \hspace*{29 mm} Berry's phase) \\ 
 \hspace*{12 mm} 03.65.Ca Formalism
\\
In this note we explore the consequences of the hypothesis of the
fundamental length in quantum mechanics motivated by 
the analyses of the noncontractible space in particle physics \cite{Pal1}
 and cosmology \cite{Pal2}. 
We first briefly discuss the formalism of Mead \cite{Mead}.
He introduced indeterminate operators in order to treat the
"fundamental length algebra" in complete analogy with the Heisenberg
algebra of quantum mechanics \cite{Mead}:
\begin{eqnarray}
  [r_{\alpha},Q_{\beta}]=\imath l \beta (R_{c}/l) \delta_{\alpha
 \beta},\ \  \alpha=1,2,3 \hspace*{40 mm} \\
 -1 \leq  Q_{\alpha} \leq 1,\ for\ spectrum\ of\ Q_{\alpha},\hspace*{35 mm}
 \nonumber \\
 \beta (y)={\cal O}(1)\ when\ y={\cal O}(1),\hspace*{30 mm} \nonumber \\
 l=fundamental\ length,\ R_{c}=characteristic\ scale\ of\ the\ physical\ 
 system. \nonumber
\end{eqnarray}
For the single-particle moving in a potential field $V(\vec r)$, 
applying " the fundamental length algebra" and the Heisenberg uncertainty
principle, he found the following uncertainty relation for Hamiltonian
\begin{eqnarray}
      [r_{\alpha},Q_{\beta}]=\imath l \delta_{\alpha,\beta}, \hspace*{50 mm}
 \nonumber \\
      H = p^{2}/2m + V(\vec{r}), \hspace*{40 mm} \nonumber \\
      \triangle H \geq \frac{1}{R_{c}}|l \langle\vec{r}\cdot\vec{\nabla} V
\rangle   
   - (1/m)\langle\vec{p}\cdot\vec{Q}\rangle| .\hspace*{15 mm}  
\end{eqnarray}
Under the assumption that no cancellation between the two terms in the bracket
takes place, one can estimate the spread of the frequency of the "broadened
states" as \cite{Mead}

\begin{eqnarray}
  \triangle \nu \geq \nu_{0}(l/R_{c}) \beta (R_{c}/l) .\hspace*{40 mm} 
\end{eqnarray}

In the following we want to show that the effect of the line broadening
can be correctly resolved by the inclusion of Berry's geometrical phase
\cite{Berry}.  

Following Kuratsuji and Iida \cite{Kurats}, let us consider the Hamiltonian
with a rotational symmetry consisting of the collective and internal parts:

\begin{eqnarray}
H = H_{0}(P_{\phi})+\left( \begin{array}{cc} R_{c} & r e^{\imath\phi} \\
r e^{-\imath\phi} & -R_{c} \end{array} \right), \hspace*{30 mm} 
\end{eqnarray}

and the corresponding Berry's phases for the upper and lower states are
 
\begin{eqnarray}
\Gamma_{\pm} = \mp \pi (1-\frac{R_{c}}{\sqrt{R_{c}^2+r^2}}). \hspace*{35 mm}
\end{eqnarray}

We suppose that the material point is now spread over and is
represented as a patch at the pole of the sphere with a  
diameter equal to the fundamental length $l$ 
(we assume that $R_{c} \gg l$). 
The semiclassical quantization condition corrected for Berry's phase gives us
the energy spectrum for the two-level system \cite{Kurats}:

\begin{eqnarray}
\frac{1}{2 \pi}\oint_{C} P_{\phi} d \phi = P_{\phi}(E)=(m-\frac{\Gamma}
{2 \pi})\hbar,\ m \epsilon Z , \nonumber \\
E_{\pm}^{sc} \simeq H_{0}(P_{\phi})\pm\sqrt{R_{c}^{2}+r^2}
\pm\frac{\hbar}{2}\frac{d H_{0}}{d P_{\phi}}\mp
\frac{\hbar}{2}\frac{R_{c}}{\sqrt{R_{c}^2+r^2}}\frac{d H_{0}}
{d P_{\phi}} .
\end{eqnarray}

Our interpretation of this formula as a formula for the spectrum of
the "broadened states", immediately gives us the relation for the
line-broadening effect 
(a collection of all states with $0 \leq r \leq l/2$ and 
the corresponding Berry's shifts in the spectrum define the broadened
states and the broadened spectrum lines):

\begin{eqnarray}
\triangle E \simeq \frac{\hbar}{16}(\frac{l}{R_{c}})^2
\frac{d H_{0}}{d P_{\phi}} .
\end{eqnarray}

One can observe that the effect is now quadratic in $\frac{l}{R_{c}}$, 
contrary to the derivation of Mead with the indeterminate operators
where the effect is linear in $\frac{l}{R_{c}}$. Evidently, possible
cancellation in eq.(2) can take place and Mead's formalism can hardly
resolve the problem, unlike the present geometrical approach that is
fully compatible with quantum mechanics 
 without any necessity to introduce a new operator algebra.

Our conclusion is valid for any quantum mechanical system if  
one can find a nonvanishing geometrical phase. As an example we also give 
the three-level system and the geometrical phase for SU(3) \cite{Khanna}.
If two levels of this system are almost degenerate, one can evaluate the
line broadening effect  \cite{Khanna} as

\begin{eqnarray}
spectrum\ of\ internal\ Hamiltonian:\ \mu_{1}=\frac{sin\theta}{\sqrt{3}}+cos\theta,\ 
 \nonumber \\ \mu_{2}=\frac{sin\theta}{\sqrt{3}}-cos\theta,\ 
\mu_{3}=\frac{-2 sin\theta}{\sqrt{3}}; \hspace*{40 mm} \nonumber \\
\Gamma=-\oint_{C} (cos^{2}\theta d \chi_{1}+sin^{2}\theta cos^{2}
 \phi d \chi_{2}); \hspace*{25 mm} \nonumber \\
 cos^{2}\theta \simeq (\frac{l}{2 R_{c}})^{2} \Longrightarrow\ 
\triangle E \simeq \frac{\hbar}{4}(\frac{l}{R_{c}})^{2}\frac{d H_{0}}
{d P_{\phi}}. \hspace*{25 mm}
\end{eqnarray}

We see that the pattern of the line broadening is also quadratic in
$\frac{l}{R_{c}}$. Because of this fact, it could be very difficult to
measure such a tiny effect even if $R_{c}\sim 10^{-13} cm$ (typical 
scale in nuclear physics) and if 
$l\sim 10^{-16} cm$, as suggested in refs. [1] and [2]. 

Although the elimination of other sources of broadening could represent
an insurmountable task, in the near future one can imagine some
successful measurements in nuclear physics or quantum optics 
 \cite{Mead}.

\end{document}